\documentclass[prd,nofootinbib]{revtex4}

\usepackage{slashed}
\usepackage{amssymb}
\usepackage{amsmath}
\usepackage{amsfonts}
\usepackage{graphicx}

\newcommand{\be}{\begin{equation}}
\newcommand{\ee}{\end{equation}}
\newcommand{\bea}{\begin{eqnarray}}
\newcommand{\eea}{\end{eqnarray}}

\begin{document}

\begin{flushleft}
KCL-PH-TH/2019-{\bf 43} \\
\end{flushleft}

\title{{\Large {\bf Quantum Anomalies, Running Vacuum and Leptogenesis: an Interplay\footnote{{\it Based on invited talks by NEM at Corfu Summer Institute 2018 ``School and Workshops on Elementary Particle Physics and Gravity'',
		(CORFU2018), 31 August - 28 September 2018, Corfu, Greece, and at ``HEP2019 Conference on recent developments in High-Energy Physics and Cosmology'', 
		April 17-20 2019, National Centre for Scientific Research ``Demokritos'', Athens, Greece.}}}}}


\author{Spyros Basilakos} 

\affiliation{Academy of Athens, Research Center for Astronomy and Applied Mathematics, Soranou Efessiou 4, 115 27 Athens, Greece; \\
National Observatory of Athens, Lofos Nymfon, 11852, Athens, Greece. \\
}

\author{Nick E. Mavromatos}

 \affiliation{King's College London, Physics Department, Theoretical Particle Physics and Cosmology Group,   Strand, London WC2R 2LS, UK. \\
        }

\author{Joan Sol\`a Peracaula}

\affiliation{Departament de F\'\i sica Qu\`antica i Astrof\'\i sica, and Institute of Cosmos Sciences (ICCUB), Univ. de Barcelona, Av. Diagonal 647 E-08028 Barcelona, Catalonia, Spain. \\
}



\begin{abstract}

We discuss a connection between gravitational-wave physics, quantum theory anomalies, right-handed (sterile) neutrinos, (spontaneous) CPT Violation and Leptogenesis, within the framework of string-inspired cosmological models. In particular, we present a scenario, according to which
(primordial) gravitational waves induce 
gravitational anomalies during inflation. This, in turn, results in the existence of an undiluted (at the exit of inflation/beginning of radiation era) bakcground of the Kalb-Ramond (KR) axion of the massless bosonic string gravitational multiplet. The latter may violate spontaneously CP and CPT symmetries, and induce leptogenesis during the radiation-dominated era in models involving right-handed neutrinos. The so-generated lepton asymmetry may then be communicated to the baryon sector by appropriate baryon-minus-lepton-number (B - L)-conserving, but 
(B + L)-violating, (sphaleron) processes in the Standard Model sector, thus leading to matter dominance over antimatter in the Universe.
In the current (approximately de Sitter) era, the KR axion background may provide a source for an axionic dark matter in the Universe, through its mixing with other axions that are abundant in string models. As an interesting byproduct of our analysis, we demonstrate that the anomalies contribute to the vacuum energy density of the Universe terms of 
 ``running-vacuum'' type, proportional to the square of the Hubble parameter, $H^2$.\\

\end{abstract}

\maketitle

\section{Introduction and Summary \label{sec:intro}}

In a previous talk, given by one of the authors (N.E.M.)~\cite{NEMcorfu2017} at the 2017 Corfu Summer Institute {\it Workshop on The Standard Model and Beyond}, a review was presented of an approach to leptogenesis~\cite{sarkarlepto,bms2} based on spontaneously broken Lorentz and CPT symmetries in string-inspired effective low-energy field theories, which include massive sterile right-handed neutrinos (RHN).  According to these scenarios, massless axions, associated with the 
Kalb-Ramond (KR)~\cite{Kalb} spin-one antisymmetric tensor field of the massless string gravitational multiplet~\cite{gsw,tseytlin,kaloper}, play an important dual r\^ole. First, they have been argued to induce Lorentz- (LV) and CPT- (CPTV) Violating  backgrounds in the fermion sector, of the type encountered in the Standard Model Extension (SME)~\cite{kostel}, which in turn  leads to leptogenesis in the radiation era~\cite{sarkarlepto,bms2}, through the (tree-level) CP-Violating decays of the RHN into Standard Model  particles in the presence of such backgrounds. 
In addition, and importantly, quantum fluctuations of the KR axion may also catalyse the (radiative) generation of Majorana RHN masses~\cite{mptorsion}, beyond the standard seesaw mechanism~\cite{seesaw}. In this way, a completely self-consistent mechanism for leptogenesis can be provided, by explaining also the microscopic origin of the RHN mass, which constitutes an essential ingredient of the process.  

Nonetheless, the approach so far suffered from the lack of a microscopic origin of the (approximately constant) KR background, whose presence is essential. The most pressing questions concern  the way this field is created in the primordial universe after inflation, and  the emergence of its slowly-moving LV and CPTV  background form during the radiation era, which is essential for the leptogenesis scenario of \cite{sarkarlepto,bms2}.

In the current talk, we will review a string-inspired cosmological scenario, developed in \cite{essay,bamasol}, which attempts to provide an answer to both of the above questions, but also goes much more beyond this, as it actually makes important links between quantum anomalies~\cite{anomalies,jackiw} and the very existence of the material world, including predictions on the dark energy and dark matter 
content of the Universe.  Specifically, the model implies (quantum) corrections due to anomalies on the vacuum energy density of a ``running vacuum type''~\cite{rvm1}, and also makes predictions on the nature of dark matter, which is considered to be (mostly) axionic.

 For our purposes in this talk, we shall concentrate only on those aspects of the model  of \cite{essay,bamasol} which are relevant to the aforementioned two questions in regards to the leptogenesis scenario of \cite{sarkarlepto}. In this respect, we mention that, in such a framework, the presence of an undiluted KR axion background at the exit from inflation is a consequence  of gravitational anomalies~\cite{anomalies,jackiw} induced by  primordial gravitational waves during inflation~\cite{stephon}. We stress that 
only gravitational degrees of freedom of the massless string multiplet (including the KR axion and inflaton fields that drive inflation) are considered in the string effective action during the inflationary era~\cite{essay,bamasol}. Then, the generation of chiral (fermionic) matter, including RHN,  at the exit from the inflationary phase leads to the cancelation of the gravitational anomalies in the radiation- and matter- dominated eras, as required for consistency (diffeomorphism invariance~\cite{jackiw}) of the quantum field theory of matter and radiation. Such a cancellation induces - during the radiation era - a bulk flow of the KR background, which scales with the temperature as $\propto T^3$. Such KR backgrounds are considered as slowly varying during the (brief) epoch of leptogenesis, as discussed in \cite{bms2}, and thus the main conclusions of the approach of \cite{sarkarlepto} remain unaffected.  

The structure of the talk is the following: in the  next section \ref{sec:torsion}, we review some important formal aspects of the approach, which are related to the association of the KR axions with ``torsion'' ~\cite{torsion} in string effective field theories~\cite{tseytlin,kaloper}. This is essential to give the reader an understanding of the formalism underlying the cosmological model of \cite{essay,bamasol}.
In section \ref{sec:infl}, we discuss the r\^ole of primordial gravitational waves in inducing gravitational anomalies during the inflationary de Sitter phase of the Universe, where only stringy gravitational degrees of freedom, including the KR field, are assumed to be present in the primordial effective action. As a result, the undiluted nature of the KR axion field at the exit of inflation is demonstrated, under certain conditions that are discussed explicitly. In the following section \ref{sec:lepto}, we discuss the behaviour of this field during the radiation era, after the cancellation of gravitational anomalies by the chiral fermionic matter generated at the end of the inflationary era. A KR axion field configuration scaling with the temperature as $T^3$
emerges as a consistent solution of the pertinent equations of motion. Such a KR-axion background 
is viewed as sufficiently slow moving~\cite{bms2}, so as to lead to RHN-induced leptogenesis via the scenario of \cite{sarkarlepto}, which we also review briefly here for completeness, together with the aforementioned KR-axion-induced dynamical mass generation for the RHN. Finally, in the concluding section \ref{sec:concl}, we discuss briefly the re-surfacing of the gravitational anomalies in the current (approximately de Sitter) era of the Universe, which affects the behaviour of the KR field in the present epoch. The latter is argued to acquire an approximately constant form qualitatively similar to that during inflation. We also speculate on the potential r\^ole of the KR axion as a source for Dark Matter in the universe, due to its potential mixing~\cite{mptorsion,bamasol}, during the matter era, with other axions that are abundant in string theory~\cite{arvanitaki}. It is important to stress~\cite{essay,bamasol} that the gravitational anomaly contributions to the vacuum energy density of the string Universe are of the so-called ``running-vacuum'' type~\cite{rvm1}, scaling with the pertinent Hubble parameter as $H^2$. This may lead to important phenomenological tests of the model~\cite{valent},  a discussion of which, however, falls beyond the scope of this talk.

 \section{Torsion and Kalb-Ramond Axions in Strings \label{sec:torsion}}
 
From a formal point of view, the important feature of the string-inspired field theory models of interest in this presentation is the r\^ole played~\cite{gsw,tseytlin,kaloper} by the KR  field  as a `totally antisymmetric torsion' of the pertinent gravitational theory~\cite{torsion}, which we now proceed to review briefly. We shall restrict ourselves to the basic features that are of primary importance for  a better understanding of the topics discussed in subsequent sections of this talk. 

To this end, we first notice that in contorted geometries, the Dirac (or Majorana - the relevant extension is straightforward) fermion action reads:
\begin{equation}\label{dirac}
S_\psi = \frac{i}{2} \int d^4 x \sqrt{-g} \Big( \overline{\psi}
\gamma^\mu \overline{\mathcal{D}}_\mu \psi  
- (\overline{\mathcal{D}}_\mu \overline{\psi} ) \gamma^\mu \psi \Big)
\end{equation}
where $\overline{\mathcal{D}}_\mu = \overline{\nabla}_\mu  + \dots $,
is the covariant derivative (with the $\dots$ denoting gauge-field connection parts, in case the fermions are charged, as in the case of ``contorted Quantum Electrodynamics'' (ctQED)) 
in the presence  of  torsion, which  is  introduced through  the
contorted spin connection: $\overline{\omega}_{a  b \mu} = \omega_{a b
  \mu} + K_{a  b \mu} $, where $K_{ab \mu}$  is the contorsion tensor.
The  latter  is  related  to  the  torsion  two-form  $\textbf{T}^a  =
\textbf{d   e}^a   +    \overline{\omega}^a   \wedge   \textbf{e}^b   $
via~\cite{torsion,kaloper}:     $K_{abc}    =     \frac{1}{2}    \Big(
\textrm{T}_{cab}  - \textrm{T}_{abc}  -  \textrm{T}_{bcd} \Big)$.  The
presence  of torsion  in the  covariant derivative  in  the 
action (\ref{dirac}) leads, apart from the standard terms in manifolds
without  torsion, to an  additional term  involving the  axial current
\be\label{axial}
J^\mu_5 \equiv \overline{\psi} \gamma^\mu \gamma^5 \psi~.
\ee
The relevant part of the action reads:
\begin{equation}\label{torsionpsi}
S_\psi \ni  - \frac{3}{4} \int d^4x \sqrt{-g} \, S_\mu \overline{\psi}
\gamma^\mu \gamma^5 \psi  = - \frac{3}{4} \int S \wedge {}^\star\! J^5  
\end{equation}
where $\textbf{S} = {}^\star\! \textbf{T}$  is the dual of \textbf{T}: $S_d
=   \frac{1}{3!}     \varepsilon^{abc}_{\quad   d}   T_{abc}$, 
where $\varepsilon^{\mu\nu\rho\sigma}$ is the covariant Levi-Civita tensor density in the presence of gravity, 
$\varepsilon^{\mu\nu\rho\sigma} =\frac{{\rm sgn}(g)}{\sqrt{-g}}\,  \epsilon^{\mu\nu\rho\sigma}$, $\varepsilon_{\mu\nu\rho\sigma} = \sqrt{-g}\,  \epsilon_{\mu\nu\rho\sigma}$,
with $\epsilon^{0123} = +1$, {\emph etc.}, totally antisymmetric in its indices.

In a  quantum gravity setting,  where one integrates over  all fields,
the torsion terms  appear as non propagating fields  and thus they can
be integrated out exactly. The authors of \cite{kaloper} have observed
though   that  the   classical  equations   of  motion   identify  the
axial-pseudovector torsion field $S_\mu$ with the axial current, since
the torsion equation yields
\begin{equation}\label{torsionec}
K_{\mu a b} = - \frac{1}{4} e^c_\mu \, \varepsilon_{a b c d} \, \overline{\psi}
\gamma_5 {\tilde \gamma}^d \psi\ .
\end{equation}
From this  it follows $\textbf{d}\,{}^\star\!\textbf{S}  = 0$, leading
to a  conserved ``torsion charge'' $Q =  \int {}^\star\!  \textbf{S}$.
To  maintain  this conservation  in  quantum  theory, they  postulated
$\textbf{d}\,{}^\star\!\textbf{S} = 0$ at the quantum level, which can
be  achieved  by  the  addition  of  judicious  counter  terms.   This
constraint, in a path-integral formulation of quantum gravity, is then
implemented  via a delta  function constraint,  $\delta (d\,{}^\star\!
\mathbf{S})$, and the latter via the well-known trick of introducing a
Lagrange multiplier  field $\Phi (x)  \equiv (3/\kappa^2)^{1/2} b(x)$, 
with $\kappa=\sqrt{8\pi {\rm G}} =1/M_{\rm Pl}$  the (four space-time dimensional) gravitational constant, and $M_{\rm Pl}$ the  reduced Planck mass.
Hence, the relevant torsion  part of the quantum-gravity path integral
would include a factor {\small
\begin{eqnarray}
 \label{qtorsion}
&&\hspace{-5mm} \mathcal{Z} \propto \int D \textbf{S} \, D b   \, \exp \Big[ i \int
    \frac{3}{4\kappa^2} \textbf{S} \wedge {}^\star\! \textbf{S} -
      \frac{3}{4} \textbf{S} \wedge {}^\star\! \textbf{J}^5  +
      \Big(\frac{3}{2\kappa^2}\Big)^{1/2} \, b \, d {}^\star\! \textbf{S}
      \Big]\nonumber \\  
&&\hspace{-5mm}=\!  \int D b  \, \exp\Big[ -i \int \frac{1}{2}
      \textbf{d} b\wedge {}^\star\! \textbf{d} b + \frac{1}{f_b}\textbf{d}b 
\wedge {}^\star\! \textbf{J}^5 + \frac{1}{2f_b^2}
    \textbf{J}^5\wedge\, ^\star \textbf{J}^5 \Big]\; ,\nonumber\\
\end{eqnarray}
\hspace{-1.5mm}}
where   
$f_b = (3\kappa^2/8)^{-1/2} $ 
and  the  non-propagating   $\textbf{S}$  field  has  been  integrated
out in the second equality of \eqref{qtorsion}. The reader  should notice that, as a  result of this integration,
the Lagrange multiplier field $b(x)$ became a dynamical (axion-like) particle, and  
the   corresponding   \emph{effective}   field   theory   contains   a
\emph{non-renormalizable} repulsive four-fermion axial current-current
interaction, characteristic of any torsionful theory~\cite{torsion}.

The torsion term, being geometrical, due to gravity, couples universally to all fermion species, including sterile neutrinos, if present in the theory.
Thus, in the context of the SM of particle physics,  the axial current (\ref{axial}) is expressed as a sum over fermion species
\be\label{axial2}
J^\mu_5 \equiv \sum_{i={\rm fermion~species}} \, \overline{\psi}_i  \gamma^\mu \gamma^5 \psi_i ~.
\ee
In theories with chiral anomalies, like the quantum electrodynamics part of SM, 
the  axial current  is not
conserved at the  quantum level, but its divergence
is obtained by the one-loop result~\cite{anomalies}:
\begin{eqnarray}
   \label{anom}
\nabla_\mu J^{5\mu} \! = \! \frac{e^2}{8\pi^2} {F}^{\mu\nu}
  \widetilde{F}_{\mu\nu}  
- \frac{1}{192\pi^2} {R}^{\mu\nu\rho\sigma} \widetilde
{R}_{\mu\nu\rho\sigma} 
\equiv G(\textbf{A}, \omega)\;,
\end{eqnarray}
with $\nabla_\mu$ the gravitational covariant derivative with respect to the torsion-free connection $\omega$. The reader is urged to take notice of the fact that only the torsion-free spin connection $\omega$ 
appears in the anomaly equation \eqref{anom}.  In our string-theory inspired context this can be achieved by the addition of proper renormalisation-group counter terms to
the  action~\cite{hull,kaloper}, which  can  convert the  anomaly from  a contorted one,
$G(\textbf{A},   \overline   \omega)$    to   $G(\textbf{A},
\omega)$.

Taking into account (\ref{anom}), we may then partially integrate  the second  term in  the exponent  of the
right-hand-side  of  the effective action (\ref{qtorsion}), so that the latter acquires the form:
\begin{equation}\label{brr}
 \mathcal{Z} \propto \int D b\ \exp\Big[ - i \int \frac{1}{2}
    \textbf{d} b\wedge {}^\star\! \textbf{d} b  - \frac{1}{f_b} b
    G(\textbf{A}, \omega)  
+ \frac{1}{2f_b^2} \textbf{J}^5 \wedge \, ^\star \textbf{J}^5 \Big]\; .
\end{equation}

In the context of our (four space-time dimensional) string-inspired models~\cite{NEMcorfu2017,sarkarlepto,bms2},
the totally antisymmetric component $S_\mu$ of the torsion 
is identified with the field strength of the spin-one antisymmetric tensor (Kalb-Ramond (KR)~\cite{Kalb}) field,
\begin{align}\label{Hdef}
H_{\mu\nu\rho} = \partial_{[\mu} B_{\nu\rho]},
 \end{align}
where  the  symbol  $[\dots   ]$  denotes  antisymmetrization  of  the
appropriate indices.
The string theory effective action depends only on $H_{\mu\nu\rho}$ as a consequence of the ``gauge symmetry''
$B_{\mu\nu} \rightarrow B_{\mu\nu} + \partial_{[\mu }\Theta_{\nu]} $ that characterises the closed string theory sector~\cite{gsw}.
It can be shown~\cite{tseytlin} that the
terms of the effective action 
up to and including quadratic order in the Regge slope parameter $\alpha^\prime$, of relevance to the low-energy (field-theory) limit of string theory, 
which involve the H-field strength,
can  be assembled in such a way
that  only    torsionful     Christoffel    symbols, $\overline{\Gamma}^\mu_{\nu\rho}$ 
appear:
$
  \label{torsionful}
\overline{\Gamma}^\mu_{\nu\rho}\ =\ \Gamma^\mu_{\nu\rho} +
\frac{\kappa}{\sqrt{3}}\, H^\mu_{\nu\rho}\ \ne\ 
\overline{\Gamma}^\mu_{\rho\nu}\;,$
where $\Gamma^\mu_{\nu\rho}  = \Gamma^\mu_{\rho\nu}$ is  the ordinary,
torsion-free, symmetric connection. The reader should bear in mind that the reduced Planck mass $M_{\rm Pl} = \kappa^{-1}$ is in general different~\cite{gsw} from the string mass scale $M_s=(\alpha^\prime)^{-1/2}$. 

There is a Bianchi identity associated with the structure of KR field strength \eqref{Hdef} in string theory, 
\begin{align}\label{bianchiI}
\varepsilon^{\mu\nu\rho\sigma}\, \partial_{\sigma}\, H_{\mu\nu\rho} =0.
\end{align}
In four space-time dimensions (upon assuming that the dilaton field is stabilised to a constant), 
a solution to \eqref{bianchiI} is given by~\cite{aben,kaloper}: 
\begin{align}\label{solH}
 H_{\mu\nu\rho} \propto \varepsilon_{\mu\nu\rho\sigma}\, \partial^\sigma b^{\rm KR}(x)~,
\end{align}
that is,
the H-field is the derivative of a KR axion-like field, $b^{\rm KR}(x)$. In view of the structure of \eqref{solH}, the field $b^{\rm KR}$ is also referred to as the ``dual'' of the $H$-three form.
The r\^ole of $b^{\rm KR}(x)$ is analogous to that played by the Lagrange multiplier pseudoscalar field $b(x)$ introduced previously (\eqref{qtorsion}) in the case of ctQED, 
as can be readily seen by implementing the Bianchi identity \eqref{bianchiI} as a constraint in the respective path-integral. After integrating out the field strength $H_{\mu\nu\rho}$ in the low-energy string effective action, the KR axion field becomes dynamical.\footnote{Actually, as we shall discuss later on in the article, in string theory, the 
field strength of the KR field \eqref{Hdef} is modified by appropriate (Chern-Simons) terms required by anomaly cancellation. This leads to a modification of the Bianchi identity \eqref{bianchiI}, whose implementation as a constraint in the path integral will lead to couplings of the KR axion $b^{\rm KR}(x)$ to the Chern Simons terms,
with important implications for the underlying physics.} In view of the torsion interpretation of $H_{\mu\nu\rho}$, its coupling to fermions is analogous to that of the torsion axial pseudovector $S^\alpha = \varepsilon^{\alpha\mu\nu\rho}\, T_{\mu\nu\rho}$ of ctQED to the axial fermion current $J_\alpha^5$, leading after integration of the torsion field, to four-fermion axial-current-current terms in the effective action. In view of the analogy with the ctQED case, therefore, 
from now on we denote the KR axion field also as $b(x)$, for notational convenience.
As we shall discuss below, the field $b(x)$ plays an important dual r\^ole: it induces leptogenesis during the radiation-dominated era of the Unvierse in models with massive RHN~\cite{sarkarlepto,bms2}, but it also provides self-consistent mechanisms for dynamical generation of the RHN mass itself~\cite{mptorsion}, which is crucial for generating a lepton asymmetry in this string Universe.

An important question arises, however, as to how this field survives inflation, so that it can fulfill the above r\^oles. This is what we proceed to discuss in the next section, before reviewing leptogenesis. Specifically, we shall argue that primordial gravitational waves during the inflationary phase of the string Universe induce gravitational anomalies, which in turn are held responsible for 
undiluted KR axion backgrounds at the end of inflation~\cite{essay,bamasol}.

\section{Primordial Gravitational Waves during Inflation and Undiluted Kalb-Ramond Backgrounds \label{sec:infl}}

The essential assumption made in~\cite{essay,bamasol} is that only gravitational degrees of freedom of the (bosonic) massless stringy gravitational multiplet, {\it i.e}. spin-2 graviton, spin-0 dilaton ($\Phi$) and the spin-1 antisymmetric tensor (KR) field, exist in the effective action $S_B$ during the inflationary phase:\footnote{In superstring theories, gravitinos (spin-3/2 supergravity partners of gravitons), as well as dilatinos (spin-1/2 supergravity partners of dilatons) are also included, but we shall not discuss them explicitly here.} 
\be\label{sea}
S_B  =\; \int d^{4}x\sqrt{-g}\Big( \dfrac{1}{2\kappa^{2}} [-R + 2\, \partial_{\mu}\Phi\, \partial^{\mu}\Phi] - \frac{1}{6\kappa^2}\, e^{-4\Phi}\, { H}_{\lambda\mu\nu}{H}^{\lambda\mu\nu}  + \dots \Big),
\ee
where  the $\dots$ represent higher derivative terms, as well as potential vacuum energy terms that arise in string theory models from bulk moduli fields. These latter contributions are ignored for our purposes here.  In our scenario we assume that the dilaton varies slowly, such that $\Phi \simeq \Phi_0$ a constant, which implies an (approximately) constant string coupling $g_s = g_s^{(0)} e^{\Phi_0}$. Without loss of generality we henceforth set $\Phi_0 = 0$.

In the presence of gauge and gravitational fields, cancelation of anomalies requires the modification of the field strength ${ H}_{\mu\nu\rho}$ by appropriate gauge (Yang-Mills) and Lorentz Chern--Simons terms~\cite{gsw}. As a {result,} one has the modified (anomalous) Bianchi identity
\begin{equation}\label{modbianchi2}
 \varepsilon_{abc}^{\;\;\;\;\;\mu}\,  \nabla_\mu \, {H}^{abc}_{\;\;\;\;\;\;} =  \frac{\alpha^\prime}{32} \, \sqrt{-g}\, \Big(R_{\mu\nu\rho\sigma}\, \widetilde R^{\mu\nu\rho\sigma} -
F_{\mu\nu}\, \widetilde F^{\mu\nu}\Big) \equiv \sqrt{-g}\, {\mathcal G}(\omega, \mathbf{A})\,.
\end{equation}
The tildes denote the dual tensors, as usual. Latin indices are tangent-space indices. The anomaly ${\mathcal G}(\omega, \mathbf{A})$  can be expressed as a total derivative, $\partial_\mu \Big(\sqrt{-g} \, {\mathcal K}^\mu (\omega) \Big)$. Implementing the constraint \eqref{modbianchi2} via a Lagrange multiplier field $b(x)$  in the path integral of the action \eqref{sea},  in a similar way to the case of the  
conserved ``torsion charge '' constraint  $\textbf{d}\,{}^\star\!\textbf{S}  = 0$, discussed in the previous section \ref{sec:intro},  and 
 integrating over the ${ H}$ field, one obtains the effective action~\cite{essay,bamasol}
\begin{align}\label{sea4}
S^{\rm eff}_B =\int d^{4}x\sqrt{-g}\Big[ -\dfrac{1}{2\kappa^{2}}\, R + \frac{1}{2}\, \partial_\mu b(x) \, \partial^\mu b(x)  -
 \sqrt{\frac{2}{3}}\,
\frac{\alpha^\prime}{96 \, \kappa} \, \partial_\mu b(x) \, {\mathcal K}^\mu + \dots \Big],
\end{align}
with $\dots$ denoting higher derivative terms. As already mentioned, in general, in string theory~\cite{gsw} $\alpha^\prime \ne  \kappa^2$. For concreteness in our approach though we take $\alpha^\prime \sim \kappa^2$. This will not affect qualitatively our conclusions. The reader should notice the CP- violating anomalous interactions of 
the axion field $b(x)$ with the  gravitational fields inside ${\mathcal K}^\mu$.  

While gravitational anomalies are absent in the Friedman Lema\^\i tre-Robertson-Walker (FLRW) space-time background, they appear when (primordial) gravitational-waves exist during the inflationary phase~\cite{stephon,essay,bamasol}.
Taking into account the CP-violating derivative coupling of the KR axion to the anomaly current ${\mathcal K}^\mu$ in \eqref{sea4},
one finds, up to second order in graviton fluctuations and to leading order in $k \, \eta \gg 1$ ~\cite{stephon}:
\begin{align}\label{rrt}
  \langle R_{\mu\nu\rho\sigma}\, \widetilde R^{\mu\nu\rho\sigma} \rangle  = \frac{16}{a^4} \, \kappa^2\, \int \frac{d^3k}{(2\pi)^3} \, \frac{H^2 }{2\,k^3} \, k^4 \, \Theta + {\rm O}(\Theta^3) ,
 \end{align}
where $\langle \dots \rangle$ denote an average over quantum graviton fluctuation in an (approximately) de-Sitter space-time during inflation, with scale factor $a = \exp(H t)$, where $H$ is the (approximately) constant Hubble parameter;  $k$ is a Fourier mode,  and the $k$-integral in \eqref{rrt} is restricted to physical modes $k/a$ satisfying  $k\, \eta  < \mu /H,$ where $\eta=H^{-1} \exp(-Ht)$ (resp. $t$) is the conformal (cosmic) time in the de Sitter era, and $\mu$ is a UV  cutoff.
The quantity 
\begin{align}
\Theta = \sqrt{\frac{2}{3}}\, \frac{\kappa^3}{12} \, H \,  {\dot {b}} \, \ll \, 1~,
\end{align}
with the  dot denoting, as usual, the derivative $d/dt$, arises from the anomalous CP violating coupling of the KR axion with the gravitational anomaly in \eqref{sea4}.

 The crucial point to note is that the presence of a non-trivial gravitational anomaly  \eqref{rrt} leads to a violation of  general covariance. As noted in \cite{essay,bamasol}, this should not be viewed as a `catastrophe' in the absence of matter, as the associated non-conservation of the $b$-axion stress tensor expresses an exchange of energy among the (quantum) gravitational degrees of freedom of the string multiplet.
 
From (\ref{sea4}), the classical field equation of  $b(x)$ leads, for homogeneous and isotropic space-times:
\begin{align}\label{krbeom2}
\partial_{\alpha}\Big[\sqrt{-g}\Big(\partial^{\alpha}\bar{b}  -  \sqrt{\frac{2}{3}}\,
\frac{\alpha^\prime}{96 \, \kappa} \, {\mathcal K}^{\alpha}(t)  \Big)\Big] = 0 \quad \Rightarrow  \quad \dot{\overline{b}}  =  \sqrt{\frac{2}{3}}\, \frac{\alpha^\prime}{96 \, \kappa} \, {\mathcal K}^{0}\sim {\rm constant}\,,
\end{align}
thus implying the existence of a background solution $\overline b (t)$ that respects the (large scale) homogeneity and isotropy,
with a slow-roll evolution for the gravitational anomaly~\cite{essay,bamasol}:
\begin{eqnarray}\label{k02}
{\mathcal K}^0 (t)
 \simeq {\mathcal K}^0_{\rm begin} (0) \, \exp\Big[  - 3H\, t \, \Big( 1  -  0.73 \,  \times 10^{-4} \,  \Big(\frac{H}{M_{\rm Pl}}\Big)^2 \, \Big(\frac{\mu}{M_{\rm Pl}}\Big)^4 \Big)\Big].
\end{eqnarray}
The slow-roll is a consequence of the  CMB measurements~\cite{planck}, which during inflation imply  $\frac{H}{M_{\rm Pl}}\,  \in \, \Big[ 10^{-5} , 10^{-4} \Big)$.  However,  for
$\frac{H}{M_{\rm Pl}} = \Big(10.8062 \, \frac{M_{\rm Pl}}{\mu}\Big)^2$, the  exponent in \eqref{k02} {\it vanishes} and, thus, the temporal component of the anomaly current, as well as the KR background ({\it cf.} \eqref{krbeom2}), are {\it approximately constant} during inflation, hence {\it undiluted} at its exit.  For the phenomenologically acceptable range of $H \ll M_{\rm Pl}$~\cite{planck}, {\it transplankian modes} ($\mu\gg  M_{\rm Pl}$) should  thus be involved.

To estimate  ${\mathcal K}_{\rm begin}^0(t=0)$, we assume~\cite{essay,bamasol} that the rates of the KR-axion and of the inflaton $\varphi$  are of the same order of magnitude, the only constraint being ${\dot {\overline b}}\ll{\mathcal U}(\varphi)$, with ${\mathcal U}(\varphi)$ the inflaton potential,  so as not to upset the slow-roll condition:  $\epsilon = \frac{1}{2} \frac{1}{(H M_{\rm Pl})^2}\, {\dot \varphi}^2 \sim \frac{1}{2} \frac{1}{(H M_{\rm Pl})^2}\, {\dot {\overline b}}^2 \sim 10^{-2}$\,~\cite{planck}.~\footnote{Apart from its phenomenological relevance, 
which will become clear later on, the assumption on the equality (in order of magnitude) of the rates of the axion $b$ and inflaton $\varphi$ fields during the slow-roll phase may characterise concrete models (e.g. string or supergravity inspired inflationary models~\cite{stephon}) in which the axion is the imaginary component of a complex scalar field, with its real part corresponding to the dilaton/inflaton. In our case, since both the dilaton and the KR axion stem from the same massless gravitational multiplet of the string, such an assumption is also reasonable, assuming again that inflation is driven by the dilaton through some appropriate potential generated by string loops.} In this case one obtains
 \begin{align}\label{slowrollepsi}
 {\dot {\overline b}} \sim  \sqrt{2\,\epsilon} \, M_{\rm Pl} \, H \sim  0.14 \, M_{\rm Pl} \, H~,
 \end{align}
 providing an estimate of the {\it undiluted} KR background at the end of the inflationary period. 
 
Using \eqref{krbeom2}, then, we find  ${\mathcal K}^0 \sim {\mathcal K}_{\rm begin} (t=0) \sim 16.6 \, H \, M_{\rm Pl}^2$. 
With the help of the $b$-field stress tensor and  \eqref{slowrollepsi},  then, the anomaly contribution to the energy density of the primordial string Universe is of the form of a {\it running vacuum} correction~\cite{rvm1}:
 \begin{align}\label{enpressphib2}
 \rho^{\varphi+b} \simeq 3M_{\rm Pl}^4 \Big[3.33  \times 10^{-3} \, \Big(\frac{H_{\rm infl}}{M_{\rm Pl}}\Big)^2  + \frac{{\mathcal U}(\varphi)}{3M_{\rm Pl}^4}\Big].
  \end{align}
Inflation occurs as long as ${\mathcal U}(\varphi) \gg 10^{-2} \, (H_{\rm infl}\, M_{\rm Pl})^2 $.

The reader should notice that the (approximately constant) background \eqref{slowrollepsi} violates {\it spontnaeously} 
 Lorentz and CPT invariance. This will play a r\^ole in leptogenesis during the radiation era, as we now proceed to discuss.

\section{Kalb-Ramond Axions, Leptogenesis and Sterile-Neutrino Mass Generation \label{sec:lepto}} 

After the exit from inflation, gravitational anomalies could jeopardise the incipient radiation phase, due to the associated violation of general covariance in the quantum matter/radiation theory~\cite{jackiw}. The presence of the undiluted KR-axion rectifies this~\cite{essay,bamasol}.  First, it allows to erase the offending gravitational anomaly, and  second,  it
provides the anomalous coupling  to chiral fermionic matter, essential for leptogenesis~\cite{sarkarlepto,bms2}.  Indeed, the
generation of chiral fermionic matter, induces axial current $J^{5\, \mu}$-dependent terms in the effective action, which 
now reads
\begin{align}\label{sea6}
S^{\rm eff} =&\; \int d^{4}x\sqrt{-g}\Big[ -\dfrac{1}{2\kappa^{2}}\, R + \frac{1}{2}\, \partial_\mu b \, \partial^\mu b +  \kappa\, \,  b(x)  \, \nabla_\mu \Big(\sqrt{\frac{2}{3}}\,
\frac{1}{96} \, {\mathcal K}^\mu - \sqrt{\frac{3}{8}} \,  J^{5 \, \mu}\Big)
\Big] + \dots,
\end{align}
where the $\dots$ include fermion kinetic terms, as well as radiation field dependent terms. and the chiral fermionic current is $J^{5 \, \mu} = \sum_{j} \, \bar{\psi}_j\,  \gamma^{\mu} \,\gamma^{5}\psi_j$, with $j$ a fermion species index, including RHN. Its divergence may just {\it cancel}  the offending gravitational  anomaly ${\mathcal K}^\mu$ in \eqref{sea6} and  restore  general covariance at quantum level,  leaving possibly a (harmless from this point of view) U(1) chiral anomaly part:
\begin{align}\label{cancel}
\nabla_\mu \Big(\sqrt{\frac{3}{8}} \,  {J}^{5\, \mu} - \sqrt{\frac{2}{3}}\, \frac{1}{96} \, {\mathcal K}^\mu \Big) = { ``{\rm chiral~U(1)~anomalies}''}.
\end{align}
The  post inflationary equation of motion for the KR axion, after the cancellation of the gravitational anomalies, 
$(1/\sqrt{-g})\, \partial_\mu \Big(\sqrt{-g} \, \partial^\mu b(x) \Big) = \nabla_\mu \partial^\mu b(x)  =  {``{\rm chiral~U(1)~anomalies''}}$,   results in an axion  flow~\cite{essay,bamasol}  
\begin{align}\label{axion}
\dot{\overline b}  \propto T^3 + {\rm subleading~}(\sim T^2)~{\rm chiral~U(1)~anomaly~terms}~,
\end{align} 
varying slowly with temperature $T$ during the brief teptogenesis era~\cite{bms2}. In arriving at \eqref{axion} one took into account the scaling $a(t) \sim T^{-3}$ of the scale factor with the temperature $T$ during radiation dominance, The background \eqref{axion} is matched with \eqref{slowrollepsi} at the exit from inflation, where the temperature $T \sim H/(2\pi)$ is the Gibbons-Hawking temperature of a de Sitter observer~\cite{gh}.

Let us now discuss how such a background can induce leptogenesis. In the approach of \cite{sarkarlepto,bms2}, background space-time geometries with (approximately) constant background $H_{ijk}$ torsion (corresponding to a linear in cosmic time KR axion), where Latin indices denote spatial components, have been considered. In such cases, the H-torsion background constitutes an extra source of CP violation, necessary for leptogenesis~\cite{sarkarlepto}. 
In particular, we have considered lepton-number asymmetry originating from tree-level decays of heavy sterile (right-handed, Majorana) neutrinos (RHN)  into Standard Model (SM) leptons. The relevant part of the Lagrangian is given by: 
  \be
\label{smelag}
\mathcal{L}= {\mathcal L}_{\rm SM} + i\overline{N}\slashed{\partial}N-\frac{m_N}{2}(\overline{N^{c}}N+\overline{N}N^{c})-\overline{N}\slashed{B}\gamma^{5}N-\sum_f \, y_{f}\overline{L}_{f}\tilde{\phi}N+h.c.
\ee
where ${\mathcal L}_{\rm SM}$ denotes the SM Lagrangian,  
$N$ is the RHN field, of (Majorana) mass $m_N$,  $\tilde \phi$ is the SU(2) adjoint ($\tilde{\phi}_i=\varepsilon_{ij}\phi_j~, \, i,j=1,2,$ SU(2) indices) of the Higgs field  $\phi$, 
 and $L_{f}$ is a lepton (doublet) field of the SM sector, with $f$ a generation index, $f=e, \mu, \tau$, in a standard notation for the three SM generations; $y_f$ is a Yukawa coupling, which is non-zero and provides a non-trivial (``Higgs portal'') interaction between the RHN and the SM sectors. In the models of \cite{sarkarlepto,bms2} a single sterile neutrino species 
suffices to generate phenomenologically relevant lepton asymmetry, and hence from now on 
we restrict ourselves to the first generation ($f=e$, setting $y_e = y$). The quantity 
\begin{align}\label{background}
B_\mu = M_{\rm Pl}^{-1} \, \dot{\overline b}\, \delta_{\mu0},
\end{align} 
denotes the Lorentz, CP and CPT Violating (CPTV) background, coupling the anomalous axial current to the KR axion \eqref{sea6}. 

It is important to stress at this point that, because of the derivative form of the background \eqref{background}, we need to ensure the non-conservation of the RHN current, 
via a non-zero covariant divergence, which can lead, by partial integration, to the non trivial coupling of the $B_0$ background with RHN in \eqref{smelag}.
Due to the massive nature of the RHN, their classical axial current is not conserved, independent of any anomalies. However, if the RHN were initially massless, generating their mass dynamically, through e.g. the mechanism proposed in \cite{mptorsion}, 
then assuming {\it no}  U(1) gauge couplings for the RHN, one observes that the {\it gravitational anomalies} (second term in the middle equality of \eqref{anom}), due to graviton quantum fluctuations in the radiation era, would produce anomalous contributions to the RHN-current four divergence. 

A non conserved RHN current
leads to the non-trivial $B_0$-RHN interactions and thus 
the associated leptogenesis effects~\cite{sarkarlepto,bms2}, which we now proceed to describe briefly.
In the presence of an (approximately constant) background \eqref{background}, 
the Lagrangian (\ref{smelag}) assumes the form of a Standard Model Extension (SME) Lagrangian in a Lorentz and CPTV background~\cite{kostel}. 
A lepton asymmetry is then generated due to the CP and CPT Violating tree-level decays of the RHN $N$ into SM leptons~\cite{sarkarlepto,bms2}:
\begin{eqnarray}\label{4channels}
{\rm Channel ~I}&:& \qquad  N \rightarrow l^{-}h^{+}~, ~ \nu \, h^{0}~,  \\ \nonumber 
{\rm Channel ~II}&:& \qquad  N \rightarrow l^{+}h^{-}~,~  \overline \nu \, h^{0}~.
\end{eqnarray}
where $\ell^\pm$ are charged leptons, $\nu$ ($\overline \nu$) are light, ``active'', neutrinos (antineutrinos) in the SM sector,
$h^0$ is the neutral Higgs field, and 
 $h^\pm$ are the charged Higgs fields, which, at high temperatures, above the spontaneous electroweak symmetry breaking, of interest in this scenario, do not decouple from the physical spectrum.  As a result of the non-trivial $B_0 \ne 0$ background (\ref{background}), the decay rates of the Majorana RHN between the channels I and II are different, resulting in a Lepton asymmetry~\cite{bms2}, 
 \begin{align}\label{lepto}
 \frac{ \Delta L^{TOT}(T=T_D)}{s} \sim   \dfrac{B_{0}}{m_{N}}~, 
 \end{align}
 where $s$ is the entropy density of the Universe, and $T_D$ denotes the temperature at which 
 this asymmetry freezes out (`freezeout point'), that is when the total decay width $\Gamma$ for the decays \eqref{4channels} equals the Hubble rate of the Universe, $H (T_D) \simeq \Gamma $.

The lepton asymmetry \eqref{lepto}  can then be communicated to the baryon sector via Baryon-minus-Lepton-number (B - L) conserving (but (B + L)-Violating) sphaleron processes in the SM~\cite{krs}, thus producing the observed amount of baryon asymmetry (baryogenesis)  in the Universe of ${\mathcal O}(8 \times 10^{-11})$, as indicated by (cosmological) observations~\cite{planck}. 
By requiring such phenomenological values for the lepton asymmetry \eqref{lepto}, then, one can  
infer that 
$T_D \simeq m_N \simeq  10^5 \, {\rm GeV}$ and $B_0 ={\mathcal O}(1~{\rm MeV})$ in the case of approximately constant $B_0$~\cite{sarkarlepto}. 

The reader should bear in mind that, in the semi-analytic method of \cite{bms2}, only the following combination of parameters, involving $m_N$, enters the series expansions of the solutions 
about a point $x = m_N/T$ used to approach (via Pad\'e approximants) the freezout point $x_D \simeq 1$:
\begin{align}\label{quant}
{\mathcal I} \equiv {y}^2 \frac{M_{\rm Pl}}{ m_N}.
\end{align}
We now notice that the ratio $y^2/m_N$ appears in the expression for the SM active neutrino $\nu$ masses in the seesaw mechanism~\cite{seesaw},\footnote{One needs more than one flavours for heavy neutrinos in that case, which can be easily accommodated in the framework of \cite{sarkarlepto,bms2}.}
\begin{align}\label{active}
m_\nu \sim |y|^2 v^2/m_N.
\end{align}  
In \cite{bms2},  where a constant $B_0$ was considered, the Yukawa coupling $y \sim 10^{-5}$ and $m_N \sim 10^5$~GeV~\cite{sarkarlepto,bms2} 
gave phenomenologically relevant values for  $m_\nu$. Such parameters correspond to (cf. \eqref{quant})
${\mathcal I} \sim 10^{3}$, 
which is kept  in the extension of the analysis to $T$-dependent backgrounds \eqref{axion}, $B_0(T) \sim T^3$~\cite{bms2}, thus ensuring consistency with active neutrino phenomenology.  It is important to remark that even a background that scales with the temperature as $T^3$ can be considered as slowly varying during the (brief) era of leptogenesis~\cite{bms2}, and thus the considerations and conclusions of \cite{sarkarlepto} can be straightforwardly extended  to this case. In such a case $B_0$ is found to be of order keV, with the decoupling temperature $T_D \simeq m_N \sim O(10^6 - 10^7)$~GeV~\cite{bamasol}.
Today of course, any torsion background should be strongly suppressed, due to the lack of any experimental evidence for it~\cite{kostel}. Detailed scenarios as to how such cosmologies can evolve so as to guarantee the absence of any appreciable traces of torsion today can be found in \cite{sarkarlepto,bms2,bamasol}. We shall briefly discuss this latter issue in 
the concluding section \ref{sec:concl} of the talk. 

In addition to the above features, associated with KR backgrounds,  in ref.~\cite{mptorsion}, the effects of the quantum fluctuations of such a KR $H$-torsion, which survive the absence of any torsion background, have been considered in connection with a self consistent mechanism for the (radiative) generation of a mass for the RHN, which play a crucial r\^ole in the leptogenesis scenario of \cite{sarkarlepto,bms2}.
An important aspect of the coupling  of the $H$-torsion (or, equivalently, the KR axion quantum field $b(x)$) to
the  fermionic   matter  discussed   above  is  its   shift  symmetry,
characteristic of an axion field. Indeed, by shifting the field $b(x)$
by  a constant:  $b(x)  \to b(x)  +  c$, the  action (\ref{brr})  only
changes by  total derivative terms, such  as $c\, R^{\mu\nu\rho\sigma}
\widetilde{R}_{\mu\nu\rho\sigma}$               
and $c\, F^{\mu\nu}\widetilde{F}_{\mu\nu}$.  These terms are irrelevant for the
equations  of motion and  the induced  quantum dynamics,  provided the
fields fall off sufficiently fast  to zero at space-time infinity. 
The scenario for  the anomalous  Majorana mass generation  through torsion proposed in \cite{mptorsion}
consists of augmenting the  effective action (\ref{brr}) by terms that
break such a shift symmetry utilising kinetic mixing of $b(x)$ with other axion fields $a(x)$ that are abundant in string models, where they are provided  by  the  string
moduli~\cite{arvanitaki}. The axions $a(x)$ couple to the Majorana neutrinos with chirality-changing Yukawa-type interaction terms, with couplings $\tilde y_a$ (not to be confused with the couplings $y_f$ in the ``Higgs portal'' of the RHN model \eqref{smelag}), which could be generated by instanton effects. We shall not discuss this further here, but we refer the interested reader to \cite{mptorsion,NEMcorfu2017} for details and the associated phenomenology.

\begin{figure}[t]
 \centering
  \includegraphics[clip,width=0.40\textwidth,height=0.15\textheight]{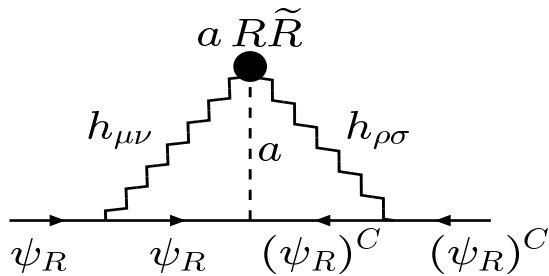} 
\caption{\it Three-loop Feynman diagram, giving rise to anomalous Majorana fermion
  mass generation~\cite{mptorsion}.  The black circle denotes the gravitationally anomalous operator $a(x)\,
  R_{\mu\nu\lambda\rho}\widetilde{R}^{\mu\nu\lambda\rho}$ induced by KR
  torsion in low-energy string effective field theories. The fields $h_{\mu\nu}$ (wavy lines) denote graviton fluctuations. Straight lines with arrows denote right handed fermion (e.g. neutrino) fields and their conjugates.}\label{fig:feyn}
\end{figure}

We only mention, for completeness, that the mechanism for  the anomalous Majorana mass generation  is shown in
Fig.~\ref{fig:feyn}.  
 Adopting  the   effective   field-theory  framework   of
\cite{dono},  the gravitationally induced Majorana mass for right-handed neutrinos, $M_R$, is estimated to be~\cite{mptorsion}:
\begin{equation}
  \label{MR}
M_R \sim \frac{1}{(16\pi^2)^2}\;
\frac{\tilde y_a\, \gamma\  \kappa^4 \Lambda^6}{192\pi^2 f_b (1 - \gamma^2 )},\end{equation} 
where $|\gamma| < 1$ is the kinetic mixing parameter and $\Lambda$ is the ultra-violet (UV) cutoff. 
In a UV
complete theory  such as  strings, the cutoff  and the Planck mass scale $M_P = \kappa^{-1}$  are related.
On using $f_b = (3\kappa^2/8)^{-1/2} = \frac{M_P}{\sqrt{3\pi}}$, $M_P \sim 2.4 \times 10^{18}$~GeV, and assuming $|\gamma | \ll 1$, 
we may approximate  \eqref{MR} as 
\begin{equation}
  \label{MRest}
M_R \sim (3.1\times 10^{11}~{\rm GeV})\bigg(\frac{\tilde y_a}{10^{-3}}\bigg)\;
\bigg(\frac{\gamma}{10^{-1}}\bigg)  
\bigg(\frac{\Lambda}{2.4 \times 10^{18}~{\rm GeV}}\bigg)^6\, .
\end{equation}
from which, for $\Lambda \sim M_P$, we observe that 
one may obtain the RHN masses in the range $m_N \sim 10^5 - 10^7$ GeV considered in the leptogenesis scenarios described above~\cite{sarkarlepto,bms2,essay,bamasol}, 
provided $| \tilde y_a \, \gamma | \sim 10^{-10} - 10^{-7}$, respectively.

In multi-axion $a_i(x), \, i=1,2, \dots n$ scenarios, as is the case of string theory~\cite{arvanitaki},  with mixing mass terms 
 $\delta M^2_{i,i+1}$, constrained to be
$\delta M^2_{i,i+1} < M_i M_{i+1}$,  so as to obtain a stable positive
mass   spectrum   for  all   axions, the corresponding anomalously  generated Majorana mass
may be estimated to be~\cite{mptorsion}
\begin{equation}
  \label{MRmix1}
M_R \sim 
\frac{\sqrt{3}\, y_a\, \gamma\,  \kappa^5 \Lambda^{6-2n} 
(\delta M^2_a)^n}{49152\sqrt{8}\, 
\pi^4 (1 - \gamma^2 )}\; ,
\end{equation}
for $n \leq 3$, and 
\begin{equation}
  \label{MRmix2}
M_R \sim 
\frac{\sqrt{3}\, y_a\, \gamma\,  \kappa^5 (\delta M^{2}_a)^3}{49152\sqrt{8}\, 
\pi^4 (1 - \gamma^2 )}\; \frac{(\delta
  M^{2}_a)^{n-3}}{(M^2_a)^{n-3}}\; ,
\end{equation}
for  $n >  3$.  For three axions
$a_{1,2,3}$ then, one obtains a  UV finite  (cut-off-$\Lambda$-independent) result for  $M_R$ at  the two-loop
level.

We note that in the above $n$-axion-mixing  scenarios, the anomalously
generated  Majorana mass  term depends only  on the  mass-mixing
parameters $\delta M_a^2$ of the  axion fields and not on their masses
themselves, as long as $n \le 3$.  Instead, for axion-mixing scenarios
with $n > 3$, the induced Majorana neutrino masses are proportional to
$(\delta M^2_a/M^2_a)^n$, which gives rise to an additional
suppression for heavy axions with masses $M_a \gg \delta M_a$.

\section{Outlook: Current De-Sitter Era and KR axion as a source of Dark Matter \label{sec:concl}}

In the current era, according to observations~\cite{planck}, the Universe re-enters a de Sitter phase, with a dark energy component in its energy budget starting taking over. In this phase,  
gravitational anomalies  will resurface, due to gravity-wave perturbations.  However, since the current Hubble parameter $H_0$ is much smaller than the one during inflation, the gravitational anomalies are much weaker as compared to their counterparts during inflation.  The KR-axion may exhibit~\cite{essay,bamasol} a similar slow-roll behaviour as in the inflationary era \eqref{slowrollepsi}, with running-vacuum-type $H_0^2$-contributions to the  vacuum energy density (\ref{enpressphib2}). We thus expect 
 \begin{equation}\label{bdot0}
{\dot b}_{\rm today} \sim  \sqrt{2\,\epsilon^\prime} \, M_{\rm Pl} \, H_0~.
 \end{equation}
 In general $\epsilon^\prime \ne \epsilon$. However, there are models~\cite{essay,bamasol}, which mix the axion $b$ with the string theory axions $a_i(x)$ in non-perturbatively generated (periodic, due to stringy instanton effects) potentials $U_{b-a}$ during the matter era, which may lead to mass terms for $b(x)$. It  is then quite tempting to assume 
 \begin{align}\label{equal}
 \epsilon^\prime \sim \epsilon = {\mathcal O}(10^{-2}),
 \end{align}  
 and identify the DM in the Universe with the axion background left after inflation.

Note from \eqref{bdot0} and \eqref{slowrollepsi},
that the slow-roll parameter of $b(x)$ measures the ratio of its kinetic energy, $K_b\sim (1/2)\,\dot{b}^2$,  to the critical energy density of the Universe, $\rho_c=(M_{\rm Pl} H)^2/3$. Thus,  on assuming \eqref{equal} and estimating {that} $K_b$ is roughly one order of magnitude smaller than $U_{b-a}$ - a typical situation of other cosmological fields, such as e.g. quintessence- we can get the dark matter content ($\rho_m$) in the right ballpark ~\cite{planck}:
\begin{equation}\label{eq:UbTb}
  \Omega_m=\frac{\rho_m}{\rho_c}\simeq \frac{U_b}{\rho_c}\simeq 10\frac{K_b}{\rho_c}\simeq 10\, \epsilon^\prime \simeq 10\, \epsilon = {\cal O}(0.1)\,.
\end{equation}

At present, the above comments constitute speculative remarks rather than concrete mechanisms for mass generation for the KR axion field, which can thus be identified as one component of DM. In our scenario, the other DM components may be provided by the string theory axions, which the $b$-field mixes with, either through kinetic mixing, as 
in the model of \cite{mptorsion}, or through appropriate non-perturbatively generated periodic potentials, or both. Although there are concrete examples from string theory~\cite{example} where such mutli-axion potentials exist, most of the applications concern the r\^ole of axions as quintessence fields contributing to inflation. In our approach, we view the late-epoch potentials again as quintessence potentials, but now they drive the modern de Sitter era of the Universe. However, our approach is still phenomenological, as we lack at present a concrete microscopic mechanism within string theory leading to  \eqref{bdot0}, \eqref{equal}.

Before closing the talk, we would like to make one important remark, concerning the magnitude of the KR background \eqref{bdot0}, \eqref{equal}. Since the KR background plays the r\^ole of a Lorentz- and CPT-Violating axial background parameter in the fermion sector of the Standard Model Extension, one should check whether the stringent bounds today for such a parameter are satisfied by \eqref{bdot0}. These bounds are: $B_0 < 10^{-2}$~eV, for the temporal component, and $B_i < 10^{-22}~{\rm eV}$ for the spatial components $i=1,2,3$.~\cite{kostel}. In our cosmological case, taking into account the small velocity of the Earth's frame, where most of the precision tests of the Standard Model Extension take place, with respect to the Cosmic Microwave Background (cosmological) frame~\cite{planck}, one observes that \eqref{bdot0} comfortably satisfies both bounds. This completes our discussion.

\section*{Acknowledgements}

SB acknowledges support from
the Research Center for Astronomy of the Academy of Athens in the
context of the program  ``{\it Tracing the Cosmic Acceleration}''.The work of NEM is supported in part by STFC (UK) under the research grant ST/P000258/1.The work of JS has
been partially supported by projects  FPA2016-76005-C2-1-P (MINECO), 2017-SGR-929 (Generalitat de Catalunya) and MDM-2014-0369 (ICCUB). This work is also partially supported by the COST Association Action CA18108 ``{\it Quantum Gravity Phenomenology in the Multimessenger Approach (QG-MM)}''. NEM acknowledges a scientific associateship (``\emph{Doctor Vinculado}'') at IFIC-CSIC-Valencia University, Valencia, Spain.

\end{document}